\documentclass[twocolumn,prl,showpacs]{revtex4}

\usepackage{graphicx}
\usepackage{amssymb}
\usepackage{amsfonts}

\DeclareGraphicsExtensions{.eps}

\begin{document}
\title{Teleportation of an atomic ensemble quantum state}

\date{\today}

\author{A. Dantan$^*$}

\author{N. Treps}

\author{A. Bramati}

\author{M. Pinard}

\affiliation{Laboratoire Kastler Brossel, Universit\'{e} Pierre et
Marie Curie,\\
Case 74, 4 place Jussieu, 75252 Paris Cedex 05, France}

\begin{abstract}
We propose a protocol to achieve high fidelity quantum state
teleportation of a macroscopic atomic ensemble using a pair of
quantum-correlated atomic ensembles. We show how to prepare this
pair of ensembles using quasiperfect quantum state transfer
processes between light and atoms. Our protocol relies on optical
joint measurements of the atomic ensemble states and magnetic
feedback reconstruction.
\end{abstract}


\newcommand{\beq}{\begin{equation}}
\newcommand{\eeq}{\end{equation}}
\newcommand{\beqr}{\begin{eqnarray}}
\newcommand{\eeqr}{\end{eqnarray}}
\newcommand{\lb}[1]{\label{#1}}
\newcommand{\ct}[1]{\cite{#1}}
\newcommand{\bi}[1]{\bibitem{#1}}
\newcommand{\bk}{_{\bf k}}

\maketitle
\newpage

The realization of quantum networks involving optical fields and
atomic ensembles is one of the most promising path towards robust
long distance quantum communication and information processing
\ct{chuang,zoller}. The efficient transfer of quantum states
within that network is a key ingredient for a practical
implementation \ct{zoller}. Several continuous variable
teleportation experiments with optical fields \ct{furusawa} have
shown that continuously teleporting optical quantum states with a
high efficiency was possible. On the other hand the teleportation
of a single atom or ion quantum state was demonstrated very
recently \ct{wineland}. In this Letter we present a direct scheme
to teleport an atomic spin state in a way very similar to that
used in the teleportation protocols for optical field states
\ct{kimble}, which can hence be efficiently integrated within a
light-atom quantum network, for instance.

Because of the long lifetime of their
ground state spins atomic ensembles are good candidates to store
and manipulate quantum states of light \ct{lukin}. We base ourselves on proposals
predicting quasiperfect quantum state transfer between field and
atoms \ct{dantan1,dantan3,dantan4} and propose to achieve the
teleportation of an atomic ensemble (1) quantum state using an
Einstein-Podolsky-Rosen-correlated pair of atomic ensembles (2)
and (3). An optical joint measurement of
the unknown ensemble 1 and one of the entangled ensemble 2 is then
performed by Alice who sends the results to Bob. Using a suitable
magnetic field Bob can reconstruct the input state on the other
correlated ensemble 3. The quasiideal character of the atom-field
quantum transfer processes allows high fidelity teleportation for
easily accessible
experimental parameters.

Another atomic teleportation protocol, relying on successive
measurements alternating with optical displacements performed on
two ensembles, was proposed by Kuzmich and Polzik \ct{kuzmich}.
However, this protocol requires several exchanges of information
between Alice and Bob. Our scheme, being a direct adaptation of
the teleportation protocols for light, needs two simultaneous
measurements to achieve real-time quantum teleportation, and can
easily be extended to other quantum communication and information
protocols, such as entanglement swapping and quantum repeaters.
This article successively describes the three steps of atomic
teleportation : preparation, joint measurement and
reconstruction.\\

\begin{figure}[h]
\includegraphics[scale=0.82]{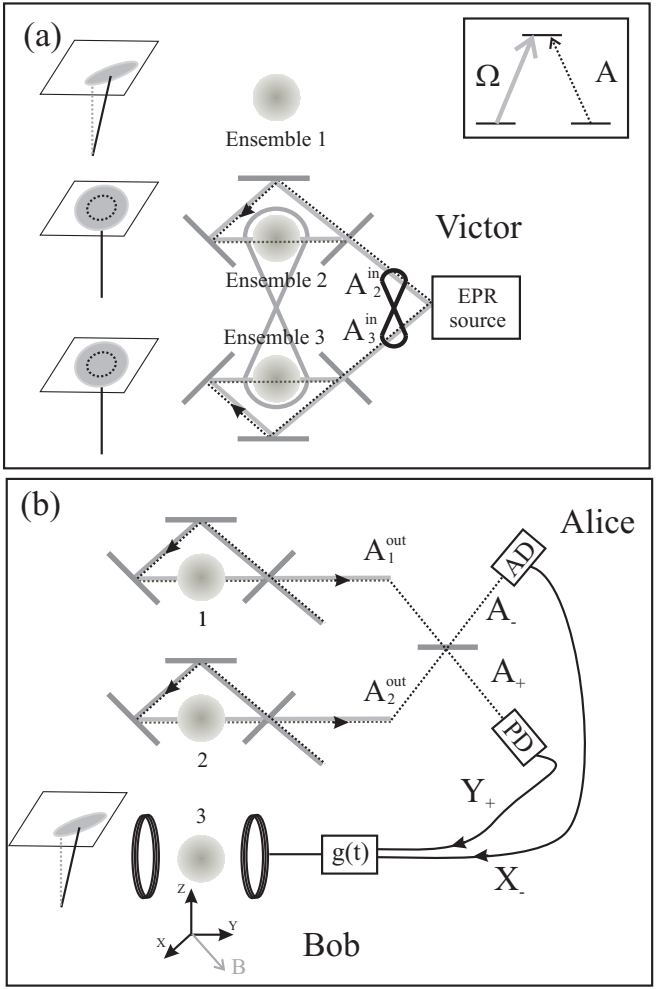}
\caption{Teleportation scheme: (a) \textit{Preparation}. Insert:
$\Lambda$-type level structure considered. Left: schematic atomic
initial states for each ensemble [for spins 2 and 3, the dashed
circle indicates the coherent spin state fluctuation
distribution]. (b) \textit{Measurement and reconstruction}. Left:
teleported state after reconstruction is completed. AD: amplitude
detector. PD: phase detector.}\label{fig1}
\end{figure}

\textit{Preparation}. We consider three $N$-atoms atomic
ensembles, labelled 1, 2, 3, with an energy level structure in
$\Lambda$ [Fig. \ref{fig1}(a)]. We assume that they are placed
inside optical cavities, for which the input-output theoretical
treatment of the atom-field quantum fluctuations is well-adapted.
They interact with a coherent control field $\Omega_i$ and with a
vacuum field $A_i$. During the preparation stage Victor pumps the
ensembles with the control fields, so that their ground state
collective spins are aligned along the $z$-axis: $\langle
J_{zi}\rangle=N/2$ ($i=1-3$). Ensemble 1 is assumed to be almost
completely spin-polarized along $z$, with a small tilt
corresponding to a non-zero coherence: $\langle
J_{z1}\rangle\simeq N/2$ and $\langle J_{x1}\rangle,\langle
J_{y1}\rangle\sim\mathcal{O}(\sqrt{N})$. This means that we
consider small planar displacements of the spin in the vicinity of
the north pole of the Bloch sphere. This approximation is all the
more correct as the number of atoms is large. The quantum state of
an ensemble is then determined by the ground state coherence, the
spin components $J_x$ and $J_y$ obeying a commutation relation
$\langle[J_x,J_y]\rangle=i\langle J_z\rangle=iN/2$, similar to
that of an harmonic oscillator. In the Gaussian approximation the
atomic quantum state can then be represented by a noise ellipsoid
in the $(x,y)$-plane orthogonal to the mean spin. It is then
completely characterized by the coherence mean value and its
variances $\Delta J_x^2$ and $\Delta J_y^2$, which are equal to
$N/4$ for a coherent spin state for instance.

Let us suppose that the atomic state to be teleported is that of
ensemble 1, prepared by Victor, unknown to Alice and Bob. With a
suitable interaction \ct{dantan1,dantan3} Victor can prepare any
Gaussian state (coherent state, squeezed state...) by an adequate
choice of the field state $A_1^{in}$, the state of which can be
perfectly mapped onto the atoms. More explicitly, the "tilt"
depends on the field intensity and phase, whereas the noise
ellipsoid is given by the field quantum fluctuations \ct{dantan3}.
Moreover, following a method detailed in \ct{dantan4}, Victor can
also entangle ensembles 2 and 3 by transferring the quantum
correlations from a pair of EPR-fields to the two spins. This can
be achieved using an "EIT" (one- and two-photon resonant) or a
"Raman" (large one-photon detuning, but two-photon resonant)
interaction between the fields and the atoms. In these two
configurations there is little to no dissipation and the quantum
fluctuations are predicted to be conserved in atom-field quantum
state transfer processes, but other situations to generate these
EPR ensembles could also be envisaged \cite{polzik}. Since the
mean spins are parallel and equal, $J_{x2}-J_{x3}$ and
$J_{y2}+J_{y3}$ are the equivalent of the usual EPR operators,
satisfying $\langle [J_{x2}-J_{x3},J_{y2}+J_{y3}]\rangle=i\langle
J_{z2}-J_{z3}\rangle=0$. We can assume without loss of generality
that the fluctuations of $J_{x2}$ and $J_{x3}$ are correlated and
those of $J_{y2}$ and $J_{y3}$ anti-correlated, so that the
condition for their inseparability reads \ct{duan}\beqr\nonumber
\Delta(J_{x2}-J_{x3})^2+\Delta(J_{y2}+J_{y3})^2<|\langle
J_{z2}\rangle|+|\langle J_{z3}\rangle|=N.\eeqr  In a symmetrical
configuration the amount of entanglement is given by the sum of
the EPR variances (normalized to 2) \ct{giedke}\beqr\label{I23}
\mathcal{I}_{2,3}=\frac{2}{N}[
\Delta(J_{x2}-J_{x3})^2+\Delta(J_{y2}+J_{y3})^2].\eeqr When the
preparation stage is over all fields are switched off, and one
disposes of an unknown atomic quantum state 1 and an
EPR-correlated pair 2 and 3.\\

\textit{Joint measurements}. Alice then performs a simultaneous
readout of ensembles 1 and 2 by rapidly switching on the control
fields in cavities 1 and 2. As shown in \ct{dantan3} the states of
spins 1 and 2 imprint in a transient manner onto the outgoing
fields exiting the cavities $A_1^{out}$ and $A_2^{out}$. These two
fields are then mixed on a 50/50 beamsplitter and Alice performs
two homodyne detections of the resulting modes
$A_{\pm}=(A_1^{out}\pm A_2^{out})/\sqrt{2}$ [Fig. \ref{fig1}(b)].
To obtain maximal information about the initial state, Alice
measures the noise of two orthogonal quadratures - say
$X_-=A_-+A_-^{\dagger}$ and $Y_+=i(A_+^{\dagger}-A_+)$ - and sends
the results to Bob who disposes of ensemble 3. As we will show
further Bob can then reconstruct state 1 using a suitable magnetic
field and achieve teleportation.

In more details, assuming one- and two-photon resonances
("EIT"-type interaction), Alice rapidly switches the control field
on in ensembles 1 and 2 at $t=0$. The outgoing modes can be
expressed as a function of the initial atomic operators in
ensembles 1 and 2 \ct{dantan3} \beqr \label{generale}
X_i^{out}(t)&=& X_i^{in}(t)-\alpha
J_{xi}(0)e^{-\tilde{\gamma}_0t}\\\nonumber &&-2\eta^2[
X_i^{in}(t)-\tilde{\gamma}_0\int_0^t
e^{-\tilde{\gamma}_0(t-s)}X_i^{in}(s)ds]
\\\nonumber &&+\beta[ X_{vi}(t)-\tilde{\gamma}_0\int_0^t
e^{-\tilde{\gamma}_0(t-s)} X_{vi}(s)ds],\eeqr ($i=1,2$), with
$\eta^2=2C/(1+2C)$, $\alpha=\eta\sqrt{8\tilde{\gamma}_0/N}$,
$\beta=2\eta/\sqrt{1+2C}$, $C$ being the cooperativity parameter
quantifying the collective strength of the atom-field coupling
\ct{dantan1}. $\tilde{\gamma}_0$ represents the effective atomic
decay rate in presence of the control field. This parameter
depends on the cooperativity and the optimal pumping rate due to
the control field \ct{dantan3}, and it is related to the duration
of the transient optical pulse carrying the atomic state out of
the cavity. $\eta$ is actually related to the efficiency of the
transfer \ct{dantan1}, and is close to unity for large values of
the cooperativity. $X_{vi}$ is a noise atomic operator accounting
for noise induced by spontaneous emission and with unity white
noise spectrum. $X_{i}^{in}$ is the amplitude quadrature of the
vacuum field incident on cavity $i$. Similar expressions hold for
the phase quadratures $Y_i$, replacing $x$'s, $X$'s by $y$'s,
$Y$'s. To derive (\ref{generale}) we assumed a cavity frequency
bandwidth much larger than $\tilde{\gamma}_0$. In (\ref{generale})
the amplitude of the term proportional to $J_{x}(0)$ shows how the
atomic state reflects on the outgoing field state. The other terms
correspond to intrinsic optical field noise ($\varpropto X^{in}$)
and added atomic noise ($\varpropto X_{v}$). The photocurrents
measured by Alice can be expressed as a sum of these noise terms
and the atomic state: \beqr i_- &\varpropto &X_-\sim noise+
[J_{x1}(0)-J_{x2}(0)]e^{-\tilde{\gamma}_0t},\\ i_+&\varpropto &
Y_+\sim noise+ [J_{y1}(0)+J_{y2}(0)]e^{-\tilde{\gamma}_0t}.\eeqr
By choosing the right temporal profile of her local oscillator it
was shown in \ct{dantan3} that Alice can measure with a great
efficiency the atomic states, which corresponds to the joint
measurements used in the continuous
variable teleportation protocols for light.\\

\textit{Reconstruction}. From Alice's results and his correlated
ensemble 3 Bob is therefore in principle able to deduce the
initial state of ensemble 1. Were we dealing with light beams Bob
could directly feed Alice's measurements to standard
phase/intensity modulators to reconstruct state 1
\ct{kimble,furusawa}. The difficulty with an atomic ensemble is to
physically implement the reconstruction stage. An all optical
method was proposed in \ct{kuzmich}. Another way to control the
quantum fluctuations of an atomic ensemble is to use a magnetic
field in order to have the spin precess in a controlled manner.
Such a method was proposed to generate spin squeezing \ct{wiseman}
and was successfully implemented recently by Geremia \textit{et
al.} to continuously monitor the atomic spin noise via feedback
\ct{geremia}. We propose here to use a transverse magnetic field,
the components of which are proportional to Alice's homodyne
detection results. Indeed, if we choose the components of the
magnetic field, $B_x$ and $B_y$, to be proportional to $-i_+$ and
$i_-$ we will couple $J_{x3}$ to $i_-$, and $J_{y3}$ to $i_+$.
Since spin 2 and 3 are initially correlated, we intuitively expect
their correlated noises to cancel leaving only spin 1 state
imprinted onto that of spin 3 at the end of the reconstruction
phase.

More quantitatively, the Hamiltonian corresponding to the unitary
transformation that Bob performs on spin 3 is simply a
$\vec{J}.\vec{B}$ coupling \beqr
H_B=-\lambda[B_x(t)J_{x3}+B_y(t)J_{y3}].\eeqr The evolution
equations of $J_{x3}$ and $J_{y3}$ are then of the form \beqr
\dot{J}_{x3}(t)=G(t) X_-(t),\hspace{0.3cm}
\dot{J}_{y3}(t)=G(t)Y_+(t),\label{dotj3x}\eeqr in which $G(t)$
gives the electronic gain of the reconstruction process. Its
temporal profile can be adjusted in order to maximize the fidelity
of the reconstruction. At this point we would like to stress that
choosing the right profile for this electronic gain is equivalent
to choosing the right local oscillator profile in Alice's homodyne
detections. We therefore choose a temporal profile in $G(t)=G
e^{-\tilde{\gamma}_0t}$ for the gain, which we know will maximize
the information that Bob gets \ct{dantan3}. After completion of
the reconstruction, i.e. for $t\gg 1/\tilde{\gamma}_0$, the final
state of $J_{x3}$, which we denote by $J_{x3}^{out}$, can be shown
to be \beqr J_{x3}^{out}=gJ_{x1}(0)+J_{x3}(0)-g
J_{x2}(0)+J_{x}^{\;noise}\eeqr in which
$g=-G\eta/\sqrt{N\tilde{\gamma}_0}$ is the normalized gain of the
teleportation protocol and $J_{x}^{\;noise}$ is a vacuum noise
operator taking into account the losses of the process. Its
explicit form is not reproduced, but it is uncorrelated with the
spin operators and its variance, which can be calculated from Eqs.
(\ref{generale}) and (\ref{dotj3x}), is related to the intrinsic
noise added during the atom-field transfer processes: $\Delta^2
J^{\;noise}_{x}=(N/2)g^2(1-\eta^2)/\eta^2$.

We assume for simplicity initial isotropic fluctuations for the
EPR-entangled ensembles, i.e. $\Delta J_{xi}^2=\Delta
J_{yi}^2=(N/4)\cosh(2r)$ ($i=2,3$), and symmetrical correlations
$\langle \delta J_{x2}\delta J_{x3}\rangle=-\langle \delta
J_{y2}\delta J_{y3}\rangle=(N/4)\sinh(2r)$. With these notations
the inseparability criterion value (\ref{I23}) is then given by
$\mathcal{I}_{2,3}=2e^{-2r}$, which is 0 for perfect EPR
entanglement and 2 for no entanglement. The normalized variance of
$J_{x3}$, after reconstruction is then \beqr\nonumber
V_{x3}^{out}&=&g^2V_{x1}+2g^2\frac{1-\eta^2}{\eta^2}\\&&+(1+g^2)\cosh(2r)-2g\sinh(2r),\eeqr
with an identical expression for the variance of $J_{y3}$. Note
that, if the gain is set to 0, one retrieves the fact that the
fluctuations of spin 3 are not modified:
$V_{x3}^{out}=\cosh(2r)=V_{x3}$. Setting a \textit{unity gain}
($g=1$), the variances of the equivalent input noises
$N_{\alpha}^{out}\equiv V_{\alpha3}^{out}-g^2V_{\alpha1}$
($\alpha=x,y$) \ct{grangier} are related to the EPR entanglement
and the losses \beqr N_{x,y}^{out}=2
e^{-2r}+2\frac{1-\eta^2}{\eta^2}.\eeqr For high entanglement
($r\gg1$) and negligible losses ($\eta\sim 1$) the equivalent
input noises go to 0, which means that the state
of spin 1 have indeed been fully teleported to spin 3.\\

At this point we can make a few comments. First, this result is
very similar to that of light beam teleportation protocols
\ct{furusawa,kimble,grangier,treps} and shows that the input noise
variances go down to 0 if Alice and Bob share perfectly entangled
ensembles ($r=\infty$) and in the absence of losses ($\eta=1$). In
absence of entanglement ($r=0$), $N_{x}^{out}=N_{y}^{out}=2$, one
retrieves the fact that two units of vacuum noise are added for
the measurement and the reconstruction in the protocol. A good
criterion to estimate the quality of the teleportation is provided
by the product of the equivalent input noise variances
$V_q\equiv\sqrt{N_x^{out}N_y^{out}}$ \ct{treps}. In the absence of
losses the classical limit of 2 is beaten as soon as one disposes
of entanglement. The equivalent input noises being independent of
the input state our teleportation protocol is unconditional. One
should note that this is true in the "small tilting" approximation
limit ; the "signal", i.e. the mean value of spin 1 ground state
coherence, has to be of the same order of magnitude as the
fluctuations of the spin: $\langle J_{x1}\rangle,\langle
J_{y1}\rangle\sim\mathcal{O}(\sqrt{N})$. Within this approximation
the various measures used in light teleportation protocols
\ct{furusawa,kimble,grangier} to assess the success of the
teleportation are valid. The non-unity gain situation can be
analyzed using T-V diagrams or other measures \ct{treps}.

Secondly, we have assumed that the measurement and the feedback
times are negligible with respect to the ground state spin
lifetime, so that ensemble 3 does not evolve before the
reconstruction. This approximation is fairly reasonable since the
ground state lifetime for cold atoms or paraffin-coated cells is
at least of the order of several milliseconds or even up to the
second \ct{kuzmich}.

Third, the intrinsic noise ($\varpropto 1/C$), that is, the noise
which does not come from the detector quantum inefficiency or
electronic noise, is expected to be rather small, thanks to the
cooperative behavior of the atoms in the cavity - $C$ can easily
be made of the order of 100-1000 using low finesse cavities. This
should ensure losses at the percent level and, therefore, a good
teleportation. High-Q cavities are not required because the
atom-field coupling is enhanced by the collective atomic behavior
($C\varpropto N$). Bad cavities are actually preferable since the
cavity bandwidth has to be much larger than the atomic spectrum
width $\tilde{\gamma}_0$.

It is also interesting to look at the physical meaning of the
magnetic reconstruction. The unity gain condition $g=1$ actually
translates into the very intuitive condition that the rotation
angle of spin 3 during reconstruction in a time
$(2\tilde{\gamma}_0)^{-1}$ should be equal to the relative spin
fluctuations: $\theta=\omega_L/(2\tilde{\gamma}_0)=1/\sqrt{N}$,
where $\omega_L$ is the Larmor frequency. This condition also
gives us the order of magnitude of the magnetic field necessary to
perform the reconstruction. For an interaction with $N=10^6$
cesium atoms on the $D_2$ line, a gyromagnetic factor of $450$
kHz/G and $\tilde{\gamma}_0=(2\pi)\; 225$ kHz, the amplitude of
the magnetic field is about $1$ mG.

Last, in order to check the quality of the teleportation Victor
can simply perform a readout of ensemble 3 with the same technique
previously used by Alice and compare the output state with the
input state that he had prepared. Another way to check that this
teleportation scheme is successful would be for Bob not to
reconstruct the atomic state, but, instead, to perform an optical
readout of ensemble 3 and use both his homodyne detection results
and Alice's results to deduce the input state. However, in this
scheme, the atomic state of 1 is never effectively teleported to
ensemble 3. The spin 1 state is actually teleported to the
outgoing field $A_3^{out}$, realizing atom-to-field teleportation.

A straightforward, but nonetheless important application of our
protocol for quantum communication is \textit{atomic entanglement
swapping}: if ensemble 1 in the previous scheme was initially
quantum correlated with another ensemble 0, the previous
teleportation scheme ensures that ensembles 0 and 3 are entangled
at the end of the process. This is of importance for the
realization of quantum networks in which quantum repeaters can
ensure good quality transmission of the quantum information over
long distances \ct{zoller}.

%%%%%%%%%%%%%%%%%%%%%%%%%%%%%%%%%%%%%%%%%%%%%%%%%%%%%%%%%%%%%%%
\begin{acknowledgments}

\end{acknowledgments}
%%%%%%%%%%%%%%%%%%%%%%%%%%%%%%%%%%%%%%%%%%%%%%%%%%%%%%%%%%%%%%%

\bigskip
$^\ast$Electronic address: dantan@spectro.jussieu.fr

%%%%%%%%%%%%%%%%%%%%%%%%%%%%%%%%%%%%%%%%%%%%%%%%%%%%%%%%%%%%%%%
\end{document}